%% file: relCTH.tex
\documentclass[envcountsame,runningheads,orivec]{llncs}
\usepackage{a4wide}
\usepackage{amsmath}
\usepackage{amssymb}
\usepackage{xspace}
\usepackage{graphicx}
\usepackage[T1]{fontenc}
\usepackage{times}
\usepackage{mathptm}
\newcommand{\IN}{\mathbb{N}}
\newcommand{\IQ}{\mathbb{Q}}
\newcommand{\IZ}{\mathbb{Z}}
\newcommand{\IR}{\mathbb{R}}
\newcommand{\calP}{\ensuremath{\mathcal{P}}\xspace}
\newcommand{\calNP}{\ensuremath{\mathcal{NP}}\xspace}
\newcommand{\calNC}{\ensuremath{\mathcal{NC}}\xspace}
\newcommand{\calPSPACE}{\ensuremath{\mathcal{PSPACE}}\xspace}
\newcommand{\calEXP}{\ensuremath{\mathcal{EXP}}\xspace}
\newcommand{\TM}{TM\xspace}
\newcommand{\CM}{CM\xspace}
\newcommand{\QM}{QM\xspace}
\newcommand{\QED}{QED\xspace}
\newcommand{\gcdex}{\operatorname{gcdex}}
\newcommand{\lcm}{\operatorname{lcm}}
\newcommand{\rem}{\operatorname{rem}}
\newcommand{\loglog}{\operatorname{loglog}}

\newcommand{\name}[1]{\textsf{#1}}
\newcommand{\COMMENTED}[1]{}

\newcommand{\mycite}[2]{\cite[\textsc{#1}]{#2}}
\newtheorem{observation}[theorem]{Observation}

\newtheorem{myexample}[theorem]{Example}
\spnewtheorem{myremark}[theorem]{Remark}{\bfseries}{\itshape}
\spnewtheorem{manifesto}[theorem]{Manifesto}{\bfseries}{\itshape}
\begin{document}
\title{Physically-Relativized Church-Turing Hypotheses}
\author{Martin Ziegler\thanks{Supported by DFG project \texttt{Zi\,1009-1/2}.
The author would also like to use this opportunity to 
express his gratitude
to \textsc{John V. Tucker} from Swansea University
for putting him on the present (scientific and career) track.}}
\institute{University of Paderborn, GERMANY}
\def\thefootnote{\fnsymbol{footnote}}\stepcounter{footnote}%
\let\temp=\labelitemi\let\labelitemi=\labelitemii\let\labelitemii=\temp
\input{mymaketitle.tex}
\maketitle
\begin{abstract}
We turn `the' Church-Turing Hypothesis from an ambiguous 
source of sensational speculations into a (collection of)
sound and well-defined scientific problem(s):

Examining recent controversies, and causes for misunderstanding,
concerning the state of the Church-Turing Hypothesis (CTH), 
suggests to study the CTH relative to
an arbitrary but specific physical theory---rather than 
vaguely referring to ``nature'' in general. 
To this end we combine (and compare) physical structuralism 
with (models of computation in) complexity theory.
The benefit of this formal framework
is illustrated by reporting on some previous,
and giving one new, example result(s) of
computability and complexity in computational physics.
\end{abstract}
\begin{minipage}[c]{0.95\textwidth}
\renewcommand{\contentsname}{}
\tableofcontents
\end{minipage}
\pagebreak

\section{Introduction}
In 1937 Alan Turing proposed, and thoroughly investigated 
the capabilities and fundamental limitations of, 
a mathematical abstraction and idealization of a computer.
This Turing machine (TM) is nowadays considered the most appropriate
model of actual digital computers, reflecting what a common
PC (say) can do or cannot, and capturing its fundamental
in-/capabilities in computability and complexity classes:
any computation problem that can in practice be solved
(efficiently) on a PC belongs to $\Delta_1$ (to $\calP$);
and vice versa. In this sense, the \TM is
widely believed to be universal; and 
problems $P\not\in\calP$, or the 
Halting problem $H\not\in\Delta_1$,
have to be faced up to as principally unsolvable 
in reality.

\subsection{Turing Universality in Computer Science and Mathematics} \label{s:CSM}
Indeed there is strong evidence for this belief:
\begin{itemize}
\item There exists a so-called \emph{universal} Turing machine (UTM),
  capable of simulating (with at most polynomial slowdown) any other
  given \TM.
\item Several other natural, yet seemingly unrelated models
  of computation have turned out as equivalent to the \TM:
 \texttt{WHILE}-programs,  $\lambda$--calculus etc.
 Notice that these correspond to real-world programming languages
 like \texttt{Lisp}!
\end{itemize}
We qualify those evidence as \emph{computer scientific}---in
contrast to the following \emph{mathematical} evidence:
\begin{itemize}
\item An integer function $f$ is \TM-computable ~iff~ it is $\mu$-recursive;
\\that is, $f$ belongs to the least class of functions 
\begin{itemize}
\item containing the constant function $0$,
\item the successor function $x\mapsto x+1$,
\item the projections $(x_1,\ldots,x_n)\mapsto x_i$,
\item and being closed under composition,
\item under primitive recursion, 
\item and under so-called $\mu$-recursion.
\end{itemize}
\end{itemize}
Observe that this is a purely (and natural, inner-) mathematical notion indeed.

\subsection{Turing Universality in Physics} \label{s:Physics}
The\footnote{To be honest, this is just \emph{one} out of a 
large variety of interpretations of this hypothesis; see, 
e.g. \cite[\textsc{Section~2.2}]{Ord}, \cite{CopelandSurvey},
or \cite{LoffCosta}} \name{Church-Turing Hypothesis}  (CTH)
claims \emph{that every function which would naturally be regarded 
as computable is computable under his \emph{[i.e. Turing's]} definition, 
i.e. by one of his machines} \cite[p.376]{Kleene52}.
Its \emph{strong} version claims
that \emph{efficient} natural computability corresponds to \emph{polynomial-time}
Turing computability.
Put differently, CTH predicts a negative answer to the following
\begin{question} \label{q:CT}
Does nature admit the existence of a system 
whose computational power strictly exceeds that of a \TM ?
\end{question}
Notice that the CTH transcends computer science;
in fact, it involves \emph{physics} as the general analysis of nature.
Hence, if the answer to Question~\ref{q:CT} turned out to be negative,
this would establish a third in addition to the above two, 
computer scientific and mathematical, 
dimensions of Turing universality (cmp. \cite{UniversalBennett,Svozil3}):
\begin{itemize}
\item The class of (efficiently) physically computable functions
  coincides with the class of (polynomial-time) Turing computable ones.
\end{itemize}
Indeed, a \TM can be built, at least in principle\footnote{%
it is for instance realized (in good approximation) by any standard PC},
and hence constitutes a physical system;
whereas a negative answer to Question~\ref{q:CT} means that,
conversely, \emph{every} physical `computer' can be simulated 
(maybe even in polynomial time) by a \TM.
Such an answer is supported by long experience in two ways:
\begin{itemize}
\item[\labelitemii] 
  the constant failure to physically solve the Halting problem ~and
\item[\labelitemii] 
  the success of simulating a plethora of physical systems on
  a \TM, \\ namely in \name{Computational Physics}.
\end{itemize}
However so far all attempts have failed to prove the CTH,
i.e. have given at best bounds on the \emph{speed} of calculations
but not on the general capabilities of computation,
based e.g. on the laws of thermodynamics \cite{Landauer,Frank}
or the speed of light (special relativity) \cite{Lloyd2}.
In fact it has been suggested that the Church-Turing Hypothesis
be included into physics as an \emph{axiom}: just like the 
impossibility of perpetual motion as a source of energy
first started as a recurring experience and was then
postulated as the Second Law of Thermodynamics.
Either way, whether axiomatizing or trying to prove
the Church-Turing Thesis, one first needs a
formalization of Question~\ref{q:CT}.

\subsection{Summary}
The CTH is the subject of a plethora of publications
and of many hot disputes and speculations.
The present work aims to put some reason into the
ongoing, and often sensational \cite{KieuCP,Seth}, discussion.
We are convinced that this requires formalizing
Question~\ref{q:CT}. However it seems unlikely to reach
consensus about one single formalization.
In fact we notice that most, if not all, disputes 
about the state of the Church-Turing Hypothesis arise
from disagreeing, and usually only implicit,
conceptions of how to formalize it. So what I propose
is a \emph{class} of formalizations, namely one for
each physical theory. 

\begin{manifesto} \label{m:Main}
\begin{enumerate}
\item[a)]
Describing the scientific 
laws of nature is the purpose and virtue of physics.
It does so by means of various physical theories $\Phi$,
each of which `covers' some part of reality (but becomes
unrealistic on another part).
\item[b)]
Consequently, instead of vaguely referring to `nature',
any claim concerning (the state of) the CTH 
should explicitly mention
the specific physical theory $\Phi$ it considers;
\item[c)]
and criticism against such a claim as `based on unrealistic presumptions' 
should be regarded as directed towards the underlying physical theory 
(and stipulate re-investigation subject to another $\Phi$, 
rather than dismissing the claim itself).
\item[d)]
Also the input/output encoding better be specified explicitly 
when referring to some ``CTH$^\Phi$'':
How is the argument $\vec x$, of natural or real numbers,
fed into the system; i.e. how does its \textsf{preparation} 
(e.g. in Quantum Mechanics) proceed operationally;
and how is the `result' to be read off
(e.g. what `question' is the system to answer)?
{\rm\mycite{Section~I}{Shipman}}
\end{enumerate}
\end{manifesto}
The central Item~b) explains for the title of the present work;
the suggestion to consider physically-relativized Church-Turing Hypotheses
``CTH$^\Phi$'' bears the spirit of the related treatment of the
famous ``$\calP=\calNP$?'' Question in \cite{relNP}.

Section~\ref{s:PT} below expands on the concept 
(and notion within the philosophy of science) of a physical theory $\Phi$
and its analogy to a model of computation in computer science.
We turn Manifesto~\ref{m:Main}b) into a research programme
(Section~\ref{s:Research}) and illustrate its benefit 
to computational physics.
Before, Section~\ref{s:PhysHypercomp} reports on previous attempts 
to disprove the CTH by examples of hypercomputers purportedly
capable of solving the Halting problem, and the respective
physical theories they exploit. 
We then significantly simplify one such example to carefully
inspect its source of computational power and, based on this
insight, are in Section~\ref{s:Constructivism} 
led to extend the above

\medskip\noindent\textbf{Manifesto \ref{m:Main} (continued).}{\it
\vspace*{-1ex}\begin{enumerate}
\item[e)]
The term ``exist'' in Question~\ref{q:CT} must be
interpreted in the sense of constructivism.
\end{enumerate}}

\section{Physical Computing}  \label{s:PhysHypercomp}
Common (necessarily informal) arguments 
in favor of the Church-Turing Hypothesis
usually proceed along the following line:
A physical system is mathematically described by
an ordinary or partial differential equation;
this can be solved numerically using time-stepping---as
long as the solution remains regular: whereas a singular
solution is unphysical anyway and/or too unstable to
be harnessed for physical computing.

On the other hand, the literature knows a variety of suggestions 
for physical systems of computational power exceeding that of a \TM;
for instance:
\begin{myexample} \label{x:Hypercomp}
\begin{itemize}
\item[i)]
  \textsf{General Relativity} might admit for space-times such that
  the clock of a \TM $M$ following one world-line seems to reach infinity
  within finite time according to the clock of an observer $O$
  starting at the same event but following another world-line;
  $O$ thus can decide whether $M$ terminates or not {\rm\cite{Etesi}}.

  However it is not known whether such space-times actually \emph{exist}
  in our universe; and if they do, how to locate them and how far off
  from earth they might be in order to be used for solving the Halting problem. 
  (Notice that the closest known Black Hole, namely 
  next to star \texttt{V\,4641}, takes at least $1600$ years to travel to).
  Finally it has been criticized that, in this approach,
  a \TM would have to actually
  run indefinitely---and use corresponding amounts of storage tape and energy.
\item[ii)]
  While `standard' quantum computers using a finite number of qubits
  can be simulated on a \TM (although possibly at exponential slowdown),
  \textsf{Quantum Mechanics} (\QM) supports operators on \emph{in}finite superpositions
  which may be exploited to solve the Halting problem {\rm\cite{Dinneen,Kieu,Calude,InfQM}}.

  On the other hand, already finite\footnote{The present world-record
  seems to provide calculations on only 28 qubits;
  and even that is rather questionable \cite{NYT}}
  quantum parallelism is in considerable
  doubt of practicality due to issues of \emph{decoherence}, i.e.
  susceptibility to external, classical noise (a kind of instability if you like); 
  hence how much more unrealistic be infinite one!
\item[iii)]
  Certain theories of \textsf{Quantum Gravitation} involve, already 
  in their mathematical formulation, combinatorial conditions which are
  known undecidable to a \TM\ {\rm\cite{Geroch}}.

  These, however, are still mere (and preliminary) theories\ldots
\item[iv)]
  A light ray passing through a finite system of mirrors
  corresponds to the computation of a Turing machine;
  and by detecting whether it finally arrives at a certain
  position, one can solve the Halting problem {\rm\cite{Tygar}}.

  The catch is that the ray must adhere to \textsf{Geometric Optics},
  i.e. have infinitely small diameter, be devoid of dispersion, and 
  propagate instantaneously; also the mirrors have to be perfect.
\item[v)]
  The above claim that singular solutions can be ruled out
  is put into question by the discovery of non-collision singularities
  in \textsf{Newtonian} many-body systems {\rm\cite{Yao,WDSmithNBody}}.

  On the other hand, the construction of these singularities 
  heavily relies on the moving particles being \emph{ideal} points
  obeying Newton's Law (with the singularity at 0)
  up to \emph{arbitrary} small distances.
\item[vi)]
  Even \textsf{Classical Mechanics} has been suggested to allow for
  physical objects which can be probed in finite time to answer
  queries ``$n\in X$'' for any fixed set $X\subseteq\IN$ 
  (and in particular for the Halting problem) {\rm\cite{Tucker}}.
\end{itemize}
\end{myexample}
Notice that each approach is based on, and in fact exploits
sometimes beyond recognition, some (more of less specific) 
physical theory. Also, the indicated reproaches against each
approach to hypercomputation
in fact aim at, and thus challenge the correctness
of, the physical theory it is based on.

\section{Physical Theories} \label{s:PT}
have been devised for thousands of years as the scientific means for
objectively describing, and predicting the behavior of, nature.
We nowadays may feel inclined to patronize e.g. \textsc{Aristotle}'s eight books,
but his concept of \emph{Elements}  (air, fire, earth, water)
constitutes an important first step towards putting some
structure into the many phenomena experienced\footnote{%
\newcounter{f:respectPT}\setcounter{f:respectPT}{\value{footnote}}
Even more, closer observation reveals that an argument like
``\emph{A rock flung up will fall down,
\emph{because} it is a rock's nature to rest on earth.}''
is no less circular than the following two more contemporary ones:
``\emph{A rock flung up will fall down,
\emph{because} there is a force pulling it towards the earth.}''
and \emph{Electrons in an atom occupy different orbits,
\emph{because} they are Fermions.}}%

Since Aristotle, a plethora of physical theories of space-time
has evolved (cf. e.g. \cite{Duhem2}), 
associated with famous names like 
\textsc{Galileo Galilei},
\textsc{Ptolemy},
\textsc{Nicolaus Copernicus}
\textsc{Johannes Kepler},
\textsc{Sir Isaac Newton},
\textsc{Hendrik Lorentz}, and
\textsc{Albert Einstein}.
Moreover theories of electricity and magnetism 
have sprung and later became unified 
(\textsc{James Clerk Maxwell})
with \textsc{Gauss}ian Optics.
And there are various\footnote{%
Remember how scientists regularly get into a fight
when starting to talk about (their conception of) 
Quantum Mechanics}\newcounter{f:ambigQM}\setcounter{f:ambigQM}{\value{footnote}}
quantum mechanical and field theories.
Then the unification process continued:
\textsf{Electricity} and \textsf{Magnetism}, been merged into \textsf{Electrodynamics},
were joined by \textsf{Quantum Mechanics} to make up \textsf{Quantumelectrodynamics} (\QED),
and then with \textsf{Weak Interaction} formed \textsf{Electroweak Interaction};
moreover \textsf{Gravitation} and \textsf{Special Relativity} became \textsf{General Relativity}.

\begin{myremark}[Analogy between a Physical Theory and a Model of Computation] 
\label{r:partialPT}
Each such theory has arisen, or rather been devised,
in order to describe with sufficient accuracy
some part of nature---while 
necessarily neglecting others. 
(Quantum Mechanics
for instance is aimed at describing elementary particles
moving considerably slower than light; whereas 
Relativity Theory focuses on very fast
yet macroscopic objects.)
We point out the analogy of a physical theory to 
a model of computation in computer science:
Here, too, the goal is to reflect some aspects
of actual computing devises 
while being unrealistic with respect to others.
(A Turing machine has unbounded working tape and hence can 
decide whether a 4GB-memory bounded PC algorithm terminates;
whereas the canonical model for 
computing devises with finite memory, a DFA
is unable to decide the correct placement of brackets.)
\end{myremark}
But what exactly \emph{is} a physical theory?
Agreement on this issue is, in addition to a means for
clearing up misunderstandings as indicated in Footnote\footnotemark[\value{f:ambigQM}],
a crucial prerequisite for treating important further questions like:
\begin{quote}
\it 
Are Newton's Laws an extension of Kepler's? {\rm\cite{Duhem}}
Does Quantum Mechanics imply Classical Mechanics---and if so, in what sense exactly?
\end{quote}
To us, such intertheory relations \cite{Intertheory,Stoelzner}
are in turn relevant in view of the above Manifesto~\ref{m:Main}
with questions as the following one:
\begin{quote}
\it Do the computational capabilities of Quantum Mechanics
include those of Classical Mechanics?
\end{quote}

\subsection{Structuralism in Physics}
Just like a physical theory is regularly obtained by
trying to infer a simple description of a family of 
empirical data points obtained from experimental measurements,
a \emph{meta}-theory of physics takes
the variety of existing physical theories
as empirical data points and tries to identify 
their common underlying structure.
Indeed the philosophy of science knows several 
meta-theories of physics, that is, conceptions 
of what a physical theory is \cite{Structuralism}:
\begin{itemize}
\item \textsc{Sneed} focuses on their mathematical aspects \cite{Sneed};
  \textsc{Stegm\"{u}ller} suggests to formalize physical theories
  in analogy to the \textsf{Bourbaki Programme} in mathematics \cite{Stegmueller,Stegmueller2}.
\item \textsc{C.F. v. Weizs\"{a}cker} envisions 
 the success of unifying previously distinct theories (recall above)
 to continue and ultimately lead into a ``\textsf{Theory of Everything}''
  \cite{Weizsaecker,Scheibe}.
 To this perspective, any other physical theory 
  (like e.g. Newton Mechanics) is merely a tentative draft \cite{Weinberg}.
\item \textsc{Mittelstaedt} emphasizes plurism in physical theories,
  that is, various theories equally appropriate to describe the
  same range of phenomena \mycite{Section~4}{Mittelstaedt}.
  Also \cite{Hager} points out (among many other things) 
  that any physical theory, or \emph{model}, is a mere
  approximation and idealization of reality.
\item \textsc{Ludwig} \cite{Ludwig} and, building thereon,
  \textsc{Schr\"{o}ter} \cite{Schroeter} propose the, for
  our purpose, most appropriate and elaborated formalization,
  based on the following (meta-)
\end{itemize}

\begin{definition}[Sketch] \label{d:PT}
A physical theory $\Phi$ consists of
\begin{itemize}
\item[\labelitemii]  
  a description of a part of nature it applies to (\textsf{WB})
\item[\labelitemii]  
  a mathematical theory as the language to describe it (\textsf{MT})
\item[\labelitemii]  
  and mapping between physical and mathematical objects (\textsf{AP}).
\end{itemize}
\end{definition}
Note that, in this setting, each physical theory has a
specific and limited range of applicability (WB): a quite
pragmatic approach, compared to the almost eschatological
conception of von Weizs\"{a}cker and Weinberg.
The only hope implicit in Definition~\ref{d:PT}, 
on the other hand, is that the variety of physical
theories keeps augmenting such as their WBs (=images of
MTs under APs) eventually 
`cover' and describe whole nature: just like a mathematical \textsf{manifold}
being covered and described by the images of Euclidean subsets 
under \textsf{charts} \cite{Milnor}.

\subsection{On the Reality of Physical Theories} \label{s:realPT}
The purpose of a physical theory $\Phi$ is to describe some part of nature.
Hence, if and when some better description $\Phi'$ is found,
a `revolution' occurs and $\Phi$ gets disposed of \cite{Kuhn}.
However this it seems to have happened \emph{lege artis} only very
rarely (and is one source of criticism against Kuhn): more commonly,
the new theory $\Phi'$ is applied to those parts of nature which
the old one would not describe (sufficiently well) while keeping
$\Phi$ for applications where it long has turned as appropriate.

\begin{myexample} \label{x:realPT}
\begin{enumerate}
\item[a)]
Classical/Continuum Mechanics (\CM) for instance is often
heard of as `wrong'---because matter is in fact composed from 
atoms circled by electrons on stable orbits---yet it still 
constitutes the theory which most mechanical engineering is 
based on. 
\item[b)]
Similarly, audio systems are successfully designed 
using Ohm's Law for (complex) electrical resistance: in spite
of Maxwell's Equations being a more accurate description of
alternating currents, not to mention \QED.
\end{enumerate}
\end{myexample}
In fact, \QM (which the reader might feel
tempted to suggest as `better', in the sense of more realistic, a theory
than \CM) has been proven to \emph{not} include or imply \CM
\cite{AxiomaticBasis}---although such claims regularly 
re-emerge particularly in popular science.
Moreover, even \QM itself is again merely\footnote{%
In particular we disagree with the, seemingly prevalent,
opinion that Quantum Theory is somehow salient
or even universal in some sense \cite{HawkingHartle,Holland}}
an approximation to 
parts of nature, unrealistic e.g. at high velocities
or in the presence of large masses.

These observations urge us to enhance Manifesto~\ref{m:Main}a+c):
\begin{manifesto} \label{m:realPT}
A physical theory $\Phi$ (like, e.g. \CM) constitutes
an ontological entity of its own: 
It \emph{exists} no less than ``points'' or ``atoms'' do.
In particular, it advisable to investigate the computational
power of, and within, such a $\Phi$
(and not dismiss it on the grounds of being unrealistic:
a tautological feature of \emph{any} theory).
Again we stress the analogy to theoretical computer science (Remark~\ref{r:partialPT})
studying the computational power of models of computation $M$ (e.g. finite
automata, nondeterministic pushdown automata, 
linear-bounded nondeterministic Turing machines:
the famous \textsf{Chomsky Hierarchy} of formal languages)
\emph{although} each such $M$ is unrealistic in \emph{some} respect.
\end{manifesto}

\section{Hypercomputation in Classical Mechanics?}
Let us exemplify Example~\ref{x:Hypercomp}vi) 
with an alternative `hypercomputer' 
similar to the one presented
in \cite{Tucker} yet stripped down to purely exhibit, 
and make accessible for further study, the core idea.

\begin{figure}\begin{center}
\includegraphics[width=0.8\textwidth]{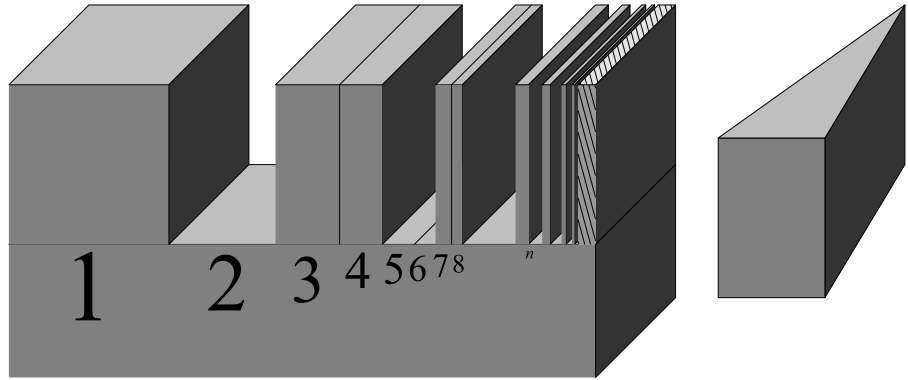} \\
\caption{\label{f:InfComb}Infinite comb with a wedge to probe its teeth}
\end{center}\end{figure}

\begin{myexample} \label{x:InfComb}
Consider a solid body, a cuboid into which has been carved
a `comb' with infinitely many teeth of decreasing width and distance,
cf. Figure~\ref{f:InfComb}. Moreover, having broken off
tooth no.$n$ iff $n\not\in H$, we arrive at an encoding
of the Halting problem into a physical object in \CM.

This very object (together with some simple mechanical control)
is a hypercomputer! Indeed it may be read off, and used to
decide for each $n\in\IN$ the question ``$n\in H$?'', 
by probing with a wedge the presence of the corresponding tooth. 
\end{myexample}

A first reproach against Example~\ref{x:InfComb} might object that
the described system, although capable of solving the Halting problem,
is no hyper\emph{computer}: because it cannot do anything else,
e.g. simulate other Turing machines. 
But this is easy to mend: just attach the system to a universal \TM,
realized in \CM \cite{Toffoli}.

The second deficiency of Example~\ref{x:InfComb} is more serious:
the concept of a solid body in \CM 
is merely an idealization of actual matter 
composed from a very large but still finite number of atoms---bad 
news for an infinite comb. 
However, as pointed out in Section~\ref{s:realPT}, 
we are to take for serious, 
and study the computational power of and within, \CM
as a physical theory.

But even then, there remains an important
\begin{observation}[Third issue about Example~\ref{x:InfComb}]
Even within \CM, i.e. granting the existence of ideal solids
and infinite combs, how are we to get hold of one encoding $H$?
Obviously one cannot \emph{construct} it from a blank
without solving the Halting problem in the first place.
Hence our only chance is to simply \emph{find} one
(e.g. left behind 
by some aliens {\rm\cite{Clarke2001,RoadsidePicknick}})
without knowing how to create one ourselves.
\end{observation}

\subsection{Existence in Physics} \label{s:Exist} 
In order to formalize the Church-Turing Hypothesis 
(a prerequisite for attempting to settle it),
we thus cannot help but notice an ambiguity about the word
``\emph{exist}'' in Question~\ref{q:CT} pointed out already
in \mycite{Remark~1.4}{InfQM}:
For a physical object to exist within a physical theory,
does that mean that
\begin{enumerate}
\item[A)] one has to
actually \emph{construct} it?
\item[B)] its \emph{non}-existence leads to a
contradiction?
\item[C)] or that its \emph{existence} 
does \emph{not} lead to a contradiction (i.e. is consistent)?
\end{enumerate}
These three opinions correspond in mathematics to the points of view
taken by a \emph{constructivist}, a \emph{classical} mathematician
(working e.g. in the \textsf{Zermelo-Fraenkel} framework), and one 
`believing' in the \textsf{Axiom of Choice}, respectively.
And at least the last standpoint (C) is well known to lead to
counter-intuitive consequences when taken in the physical realm
of \CM:
\begin{myexample}[Banach-Tarski Paradoxon] \label{x:BanachTarski}
For a solid ball (say of gold) of unit size in 3-space, there exists
a partition into finitely many (although necessarily not Lebesgue-measurable)
pieces that, when put together appropriately
(i.e. after applying certain Euclidean isometries), then form 
\emph{two} solid balls of unit size.
\end{myexample}
Note that this example is in no danger of causing inflation:
on the one hand, because actual material gold is not infinitely divisible
(cmp. the second deficiency of Example~\ref{x:InfComb});
but even \emph{within} \CM, because the partition of the ball
`exists' merely in the above Sense~C).

Hence, in order to avoid both `obviously' unnatural (counter-) 
Examples~\ref{x:InfComb} and \ref{x:BanachTarski} while sticking
to Manifesto~\ref{m:realPT}, we are led to transfer and adapt the
constructivist standpoint from mathematics to and for physics.

\subsection{Constructivism into Physical Theories!} \label{s:Constructivism}
As explained above, the ``existence'' of some physical object within a theory $\Phi$
is to be interpreted constructively.
Let us, similar to \cite{Cattaneo}, distinguish
two ways of introducing constructivism into a physical theory 
$\Phi=(\text{MT},\text{AP},\text{WB})$:
\begin{enumerate}
\item[$\alpha$)] By interpreting the mathematical theory \textsf{MT} constructively;
  compare \cite{Bishop,Kushner,Bridges,Fletcher} and \mycite{Section~III}{Stadler}.
\item[$\beta$)] By imposing constructivism onto the side of physical objects \textsf{WB}.
\end{enumerate}
It seems that Method~$\alpha$), although meritable of its own, 
does not quite meet our goal of making a \emph{physical} theory
constructive:

\begin{myexample} \label{x:EffOpen}
Consider the condition for
a function $f:X\to Y$ between normed spaces to be \emph{open}; 
or even simpler: that of the image $f[B(0,1)]\subseteq Y$ 
of the unit ball in $X$ to be an open subset of $Y$.
\begin{equation} \label{e:EffOpen}
\forall u\in f[B(0,1)] \;\;
\exists\epsilon>0\;\;
\forall y\in B(u,\epsilon) \;\;
\exists x\in B(0,1): \quad f(x)=y \enspace .
\end{equation}
A constructivist would insist that \emph{both}
existential quantifiers be interpreted constructively;
whereas in a setting of computation on
real numbers by rational approximation,
applications suffice that only $\epsilon$ be
computable from $u$, while the existence of $x$
depending on $y$ need not: compare {\rm\cite{EffOpen}}.
\end{myexample}
%
\subsubsection{Constructing Physical Objects} ~\\
The conception underlying $\beta$) is that every object in nature
(or more precisely: that part of nature described by \textsf{WB}) is 
\begin{itemize}
\item 
either a primitive one (e.g. a tree, modeled in $\Phi$
as a homogeneous cylinder of density $\rho=0.7\text{g}/\text{cm}^3$; 
or, say, some ore, modeled as \texttt{Cu\,Fe\,S}$_2$)
\item 
or the result of some technological process 
applied to such primitive objects.
\end{itemize}
The latter may for instance include 
crafting a tree into a wheel or even a wooden gear;
or smelting ore to produce bronze.

Notice also how such a process---the sequence of operations 
from cutting the tree, cleaning, sawing, carving;
or of melting, reducing, and alloying copper---constitutes 
an \textsf{algorithm} (and crucial cultural knowledge passed on 
from carpenters or redsmiths to their apprentices).
More modern and advanced science, too, knows (and teaches students)
`algorithms' for constructing physical objects:
e.g. in mechanical engineering (designing a gear, say)
or in \QM (using a furnace with boiling silver and
some magnets to create a beam of
spin-$\tfrac{1}{2}$ particles as in the famous
\textsc{Stern} and \textsc{Gerlach} Experiment
and thus operationally \emph{construct} a physical 
object corresponding via \textsf{AP} to a certain wave 
function $\psi$ as a mathematical object in \textsf{MT}).

We are thus led to extend Definition~\ref{d:PT}:
\begin{definition}[Meta-] \label{d:Meta}
The \textsf{WB} of a physical theory $\Phi$ consists of
\begin{itemize}
\item a specific collection of \emph{primitive} objects (\textsf{PrimOb})
\item and all so-called \emph{constructible} objects,
\end{itemize}
i.e., that can be obtained from primitive ones
by a sequence $(o_i)$ of \emph{preparatory operations}.
\begin{itemize}
\item
The latter are elements $o$ from a specified collection \textsf{PrepOp}.
\item
Moreover, the sequence $(o_i)$ must be ``computable''.
\end{itemize}
\end{definition}
The first two items of Definition~\ref{d:Meta}
are analogous to a mathematical theory \textsf{MT}
consisting of \emph{axioms} (i.e. claims which are true by definition)
and \emph{theorems}: claims which follow from the axioms
by a sequence of \emph{arguments}.
The last requirement in Definition~\ref{d:Meta}
is to prevent the body in Example~\ref{x:InfComb} 
from being ``constructed'' by repeated\footnote{%
Like me first,
the reader may be tempted to admit only finite sequences
of preparatory operations. However this would exclude
woodturning a handrail 
out of a wooden cylinder by letting the carving knife 
follow a curve, i.e. a continuous sequence}
``breaking off a tooth'' as preparatory operations.
On the other hand, we seem to be heading for 
a circular notion: trying to formally capture the
\emph{computational} contents of a physical theory $\Phi$
required to restrict to `constructible' objects,
which in turn are defined as the result of a
\emph{computable} sequence of preparatory operations.
That circle is avoided as follows

\medskip\noindent\textbf{Definition~\ref{d:Meta} (continued).}{\it
``Computability'' here means relative to a pre-theory $\varphi$
to, and to be specified with, $\Phi$.}

\subsubsection{Pre-Theories: Ancestry among Physical Theories} ~\\
Recall the above example from metallurgy of redoxing an ore:
this may described by the \textsf{phlogiston theory}
(an early form of theoretical chemistry, basically extending
Aristotle's concept of four \textsf{Elements} by a fifth
resembling what nowadays would be considered oxygen).
Such a `chemical' theory $\varphi$ of its own
is required to formulate (yet does not imply)
metallurgy $\Phi$, and in particular the algorithm
therein that yields to bronze: 
$\varphi$ is a \textsf{pre-theory} to $\Phi$.

We give some further, and more advanced, examples of pre-theories:
\begin{example} \label{x:prePT}
\begin{enumerate}
\item[a)]
The classical \textsf{Hall Effect} relies on \textsf{Ohm's law}
of electrical direct current
as well as on Lorentz' force law.
\item[b)]
  The Stern-Gerlach experiment, and the 
  quantum theory of spin $\Phi$ it spurred, is based on
\begin{itemize}
\item[\labelitemii]
  a classical, mechanical theory of a spinning top and precession;
\item[\labelitemii]
  some basic theory of (inhomogeneous) magnetism 
  and in particular of Lorentz force onto a dipole
\item[\labelitemii]
  an atomic theory of matter (to explain e.g. the particle beam)
\item[\labelitemii]
  and even a theory of vacuum (\textsc{Torricelli}, \textsc{von Guericke}).
\end{itemize}
\item[c)]
In fact, any quantum theory of microsystems 
requires \cite{AxiomaticBasis} 
some macroscopic pre-theory
in order to describe the devices
(furnaces, scintillators, amplifiers, counters)
for preparing and measuring the
microscopic ensembles under consideration.
\item[d)]
\textsc{Bardeen}, \textsc{Cooper}, and \textsc{Schrieffer}'s
Nobel prize-winning BCS-Theory of superconductivity
is essentially based on \QM
\item[e)]
whereas superconducting magnets, in turn,
are essential to many particle accelerators
used for exploring elementary particles.
\end{enumerate}
\end{example}
The reader is referred to \mycite{Definition~4.0.8}{Schroeter} 
for a more thorough, and formal, account of this concept.

\begin{observation} \label{o:DAG}
Technological progress can be
thought of as a directed acyclic graph:
a node $u$ corresponds to a physical theory $\Phi$;
and may be based on (one or more) predecessor nodes,
pre-theories $\varphi$ to $\Phi$.
Put differently, physical theories form
\emph{nets} or logical \emph{hierarchies}; 
cmp. 
{\rm\mycite{Vermutung~14.1.2}{Schroeter}} 
and {\rm\cite{Stoelzner}}.
\end{observation}
%

\section{Applications to Computational Physics}
Computer simulations of physical systems have over the last few decades
become (in addition to experimental, applied, and theoretical) 
an important new discipline of physics of its own. 
It has, however, received only very little support on behalf
of Theoretical Computer Science. 
Specifically, scientists working in this area
(typically highly-skilled programmers with an extensive education in,
and excellent intuition for, physics) are highly interested in,
and generally ask

\begin{question}
Why is a specific (class of) physical systems to hard
(in the sense of computing resources like CPU-time)
to simulate?
Are our algorithms optimal for them, and in what sense? 
Which are the principal limits of computer simulation?
\end{question}
Answers to such questions for various
physical systems $\Phi$ (more precisely: theories in the
sense of Section~\ref{s:PT}) are highly appreciated
in Computational Physics; answers given of course
in the language of, and using methods from,
Computational Complexity Theory \cite{Papadimitriou},
namely locating $\Phi$ in some complexity (or recursion theoretic)
class \emph{and} proving it complete for that class.

We observe that, apart from sensational attempts \cite{Seth},
there are rather few serious and rigorous 
answers to such questions to-date
\cite{Feynman,Wolfram,Moore,Shipman,Svozil,Tate,Tygar,SantaFe,Bruno}.
One reason therefor might be that, as opposed to classical 
problems considered in computational complexity, those arising
in Computational Physics
naturally involve real numbers \cite{PER,ZhongWave,ZhongSchroedinger}
where uncomputability easily occurs without completeness
\cite{Grzegorczyk,Wolfram,Moore,WDSmithNBody,CiE06}.
On the other hand, there is a well-established theory
of bit-complexity and (e.g. $\calNP$--) 
completeness over $\IR$ \cite{Friedman,Ko}
Moreover for problems \emph{defined} over real numbers
but restricted to rational inputs, the situation can become
quite subtle (and interesting): see, e.g., \cite{YapPi}
or \mycite{Proposition~30}{EffOpen}.

\subsection{Sketch of A Research Programme} \label{s:Research}
We propose a systematic exploration of the computational power
(i.e. completeness) of a large variety of physical theories.
The first goal is a general picture of physically-relativized
Church-Turing Hypotheses, that is, on 
the boundary between decidability and Turing-completeness;
later one may turn to lower complexity classes 
like \calEXP, \calPSPACE, \calNP, \calP, and \calNC.
The focus be on a thorough investigation, starting from
simplest, decidable theories and slowly proceeding towards 
more complex ones (not necessarily in historical order)
rich enough to admit a Turing- (i.e. $\Delta_2$-) complete 
system therein. In particular, it seems advisable to begin
with rather modest (rather than straight away with sexy `new')
physics:

\subsection{Celestial Mechanics}
Recall the historical progress of describing and predicting
the movement of planets and stars observed in sky
from Eudoxus/Aristotle via Ptolemy, Copernicus, and Kepler to Newton and Einstein.
Indeed, these descriptions constitute 
(not necessarily comparable, in the sense of reduction) physical theories!
The present subsection exemplifies our proposed approach
by investigating and reporting on the computational
complexities of two of them.
(We admit that, lacking any option for preparation,
celestial mechanics is of limited use as a computational system
in the sense of Manifesto~\ref{m:Main}d).

\subsubsection{Newton} ~\\
Consider a physical theory $\Phi$ of $N$ points moving in Euclidean
3-space under mutual attracting force proportional to distance$^{-2}$
(inverse-square law). This is the case for \textsf{Electrostatics}
(Coulomb) as well as for \textsf{Classical Gravitation} (Newton).
Some questions, in the sense of Manifesto~\ref{m:Main}d), may ask:
\begin{enumerate}
\item Does point \#1 reach within one second the unit ball $B$ centered at the origin?
\item Does some point eventually escape to infinity?
\item Do two points (within 1sec or ever) collide?
\end{enumerate}
It has been argued that Question~c) makes not much sense,
because a `collision' of \emph{ideal} points (recall Manifesto~\ref{m:realPT})
can be analytically continued to just pass through each other.
Note that Question~a) is not `well-posed' in case the point
just touches the boundary of $B$; it is therefore usually 
accompanied by the \emph{promise} that point \#1 either
meets the interior of $B$ within one second 
or avoids the blown-up ball $2B$ for two seconds;
and shown \calPSPACE-hard in this case \cite{Tate}.
Question~b) has only recently been shown to make sense
in that a positive answer is actually possible \cite{Xia};
and it has been shown undecidable \cite{WDSmithNBody}---however
for input configurations described by (possibly transcendental)
\emph{real} numbers given as \emph{in}finite sequences of rational approximations:
for such encodings, mere discontinuity is known to trivially imply uncomputability
without completeness \cite{Grzegorczyk}.
\subsubsection{Planar Eudoxus/Aristotle} ~\\
An early theory of celestial mechanics originates from ancient Greece.
An important purpose of it, 
and also of its successors (see Section~\ref{s:Ptolemy} below),
was to describe and predict the movement, and in particular
\textsf{conjunctions}, of planets and stars.
Let us captures this, distinguishing short-term
from long-term behavior \mycite{Section~I}{Shipman},
in the following

\begin{question} \label{q:Conjunction}
\begin{enumerate}
\item Will certain planets attain perfect conjunction, ever?
\item or within a given time interval?
\item or reach an approximate conjunction, 
  i.e. meet up to some prescribed angular distance $\epsilon$?
\end{enumerate}
\end{question}
According to \textsc{Aristotle} (Book~$\Lambda$ of \emph{Metaphysics}) 
and \textsc{Eudoxus of Cnidus}, earth resides in the center of the universe
(recall the beginning of Section~\ref{s:PT})
and is circled by \emph{celestial spheres} 
moving the celestial bodies.

\begin{figure}[htb]
\center{%
\includegraphics[width=0.6\textwidth]{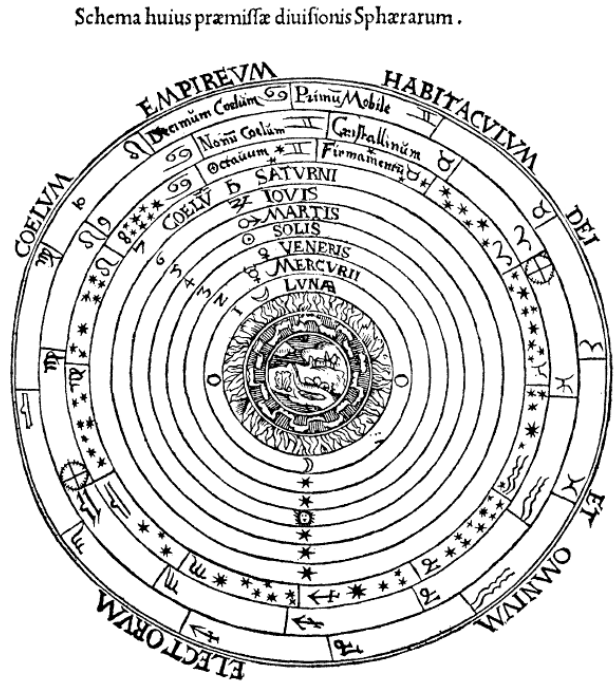}
\caption{\label{f:Ptolemy}%
Celestial orbs as drawn in \textsc{Peter Apian}'s \emph{Cosmographia} (Antwerp, 1539)}
}%
\end{figure}

\begin{definition} \label{d:Conjunction}
Let $\Phi$ denote the physical theory (which we refrain from fully
formalizing in the sense of Definition~\ref{d:PT} or even {\rm\cite{Schroeter}})
parameterized by the initial positions $u_i$ of
planets $i=1,\ldots,N$, and
their constant directions $\vec d_i$ and velocities $v_i$ of rotation.

By $\Phi'$, we mean a two-dimensionally restricted version:
planets rotate on circles perpendicular to one common direction;
compare Figure~\ref{f:Ptolemy}. Moreover, initial positions and
angular velocities are presumed `commensurable\footnote{We don't
want anybody to get drowned like, allegedly, \textsc{Hippasus of Metapontum}.
Also, since rational numbers are computable, we thus avoid
the issues from Section~\ref{s:Constructivism}.}', 
that is, rational (multiples of $\pi$).
\end{definition}
Recall that $\calNC\subseteq\calP$ 
is the class of problems solvable in
polylogarithmic parallel time on polynomially many processors;
whereas $\calP$--hard problems (w.r.t. \textsf{logspace}-reductions, say)
presumably do not admit such a beneficial parallelization.
The greatest common divisor $\gcd(a,b)$
of two given (say, $n$-bit) integers can be determined\footnote{The attentative
reader will connive our relaxed attitude concerning decision versus function problems}
in polynomial time; it is however not
known to belong to $\calNC$ nor be $\calP$-hard;
the same holds for the calculation of a
extended Euclidean representation 
``$a\cdot y+b\cdot z=\gcd(a,b)$'',
i.e. of $(y,z)=\gcdex(a,b)$ \mycite{B.5.1}{Pcomplete}.

After these preliminaries, we are able to state the computational
complexity of the above theory $\Phi'$; more precisely:
the complexity, in terms of $\Phi'$s parameters,
of the decision problems raised in Question~\ref{q:Conjunction}:

\begin{theorem} \label{t:Aristotle}
Let $k\leq n\in\IN$ and $u_1,\ldots,u_n,v_1,\ldots,v_n\in\IQ$ be given
initial positions and angular velocities (measured in multiples of $2\pi$)
of planets $\#1,\ldots,\#n$ in $\Phi'$.
\begin{enumerate}
\item[a)] Planets $\#1$ and $\#2$ will eventually appear in
  perfect conjunction ~iff~ $v_1\not=v_2\;\vee u_1=u_2$.
\item[b)] Planets $\#1$ and $\#2$ appear closer 
  than $\epsilon>0$ to each other within time interval $(a,b)$ ~iff~
  it holds, in interval notation:
\[ \emptyset \;\;\not=\;\; \IZ \;\cap\;\big( (a,b)\cdot(v_1-v_2)+u_1-u_2
  + (-\epsilon,+\epsilon)\big)
\enspace . \]
 This can be decided within $\calNC^1$.
\item[c)] The question of whether \emph{all} planets $\#1,\ldots,\#n$ 
  will ever attain a perfect conjunction, can be decided
  in $\calNC^{\gcd}$; 
\item[d)] and if so, the next time $t$ for this to
  happen can be calculated in $\calNC^{\gcdex}$.
\item[e)] Whether there exist $k$ (among the $n$) planets
  that ever attain a perfect conjunction, is $\calNP$--complete a problem.
\end{enumerate}
\end{theorem}
%

\subsubsection{Proofs} ~\\
The major ingredient is the following tool
concerning the computational complexity of 
problems about rational arithmetic progressions:

\begin{definition} \label{d:ratRem}
For $u,v\in\IQ$, let $u\div v:=(a\div b)/q$ 
and $\gcd(u,v):=\gcd(a,b)/q$ where $a,b,q\in\IZ$ are
such that $u=a/q$ and $v=b/q$ and $1=\gcd(a,b,c)$;
similarly for $u\rem v$ and $\lcm(u,v)$.

For $a,\alpha\in\IQ$, write $P_{a,\alpha}:=\{\alpha+a\cdot v:z\in\IZ\}$.
\end{definition}

\begin{lemma} \label{l:Progression}
\begin{enumerate}
\item[a)] Given $a,\alpha\in\IQ$, the unique $0\leq \alpha'<a$ with $P_{a,\alpha}=P_{a,\alpha'}$
  can be calculated as $\alpha':=\alpha\rem a$ within complexity class $\calNC^1$.
\item[b)] Given $a,\alpha$ and $b,\beta$,
 the question whether $P_{a,\alpha}\cap P_{b,\beta}=\emptyset$
 can be decided in $\calNC^{\gcd}$
\item[c)] and, if so, $c,\gamma$ with $P_{a,\alpha}\cap P_{b,\beta}=P_{c,\gamma}$
 can be calculated in $\calNC^{\gcdex}$.
\item[d)] Items~b) and c) extend from two to the intersection of $k$ 
  given arithmetic progressions.
\item[e)] Given $n$ and $a_1,\alpha_1,\ldots,a_n,\alpha_n$, determining
  the maximum number $k$ of arithmetic progressions 
  $P^{(i_1)}:=P_{a_{i_1},\alpha_{i_1}},\ldots,P^{(i_k)}:=P_{a_{i_k},\alpha_{i_k}}$ 
  that have nonempty common intersection, is $\calNP$--complete.
\end{enumerate}
\end{lemma}
A result similar to the last item has been obtained in \cite{GcdNPc}\ldots
\begin{proof}
\begin{enumerate}
\item[a)]
Notice that $P_{a,\alpha}=P_{a,\alpha'}\Leftrightarrow \alpha-\alpha'\in P_{a,0}$.
Hence there exists exactly one such $\alpha'$ in $[0,a)$,
namely $\alpha'=\alpha\rem a$. 
Moreover, integer division belongs to \calNC \cite{Beame,Chiu}.
\item[b)]
Observe that $P_{a,\alpha}\cap P_{b,\beta}\not=\emptyset$
holds ~iff~ $\gcd(a,b)$ divides $\alpha-\beta$.
Indeed, the extended Euclidean algorithm then yields
$z_1',z_2'\in\IZ$ with 
$\gcd(a,b)=-a\cdot z_1'+b\cdot z_2'$;
then $\alpha-\beta=-a\cdot z_1+b\cdot z_2$
yields $P_{a,\alpha}\ni \alpha+a\cdot z_1=\beta+b\cdot z_2\in P_{b,\beta}$.
Conversely $\alpha+a\cdot z_1=\beta+b\cdot z_2\in P_{a,\alpha}\cap P_{b,\beta}$
implies that $\alpha-\beta=-a\cdot z_1+b\cdot z_2$
is a multiple of any (and in particular the greatest)
common divisor of $a$ and $b$.
\item[c)]
Notice that $c=\lcm(a,b)=a\cdot b/\gcd(a,b)$;
and, according to the proof of b), 
$\gamma:=\alpha+a\cdot z_1$ will do, where 
$z_1,z_2\in\IZ$ with 
$\alpha-\beta=-a\cdot z_1+b\cdot z_2$
result from the extended Euclidean algorithm applied to $(a,b)$. 
\item[d)]
Notice that 
\begin{equation} \label{e:CRT1}
x\in P_{a_1,\alpha_1}\cap\cdots\cap P_{a_k,\alpha_k}
\quad\Leftrightarrow\quad 
x\equiv \alpha_i \pmod{a_i},\quad i=1,\ldots,k \enspace .
\end{equation}
According to the \textsf{Chinese Remainder Theorem},
the latter congruence admits such a solution $x$
~iff~ $\gcd(a_i,a_j)$ divides $\alpha_i-\alpha_j$
for all pairs $(i,j)$.

In order to calculate such an $x$,
notice that a straight-forward iterative 
$P_{a_{1..k-1},\alpha_{1..k-1}}\cap P_{a_k,\alpha_k}$
fails as it does not parallelize well,
and also the numbers calculated 
according to c) in may double in length
in each of the $k$ steps.
Instead, combine the $P_{a_i,\alpha_i}$ in a binary
way first two tuples $P_{a_{2j,2j+1},\alpha_{2j,2j+1}}$
of adjacent ones, then on to quadruples and so
on. At logarithmic depth (=parallel time),
this yields the desired result $x=:\alpha_0$
and $a_0:=\lcm(a_1,\ldots,a_k)$ satisfying
$P_{a_0,\alpha_0}=\bigcap_{i=1}^k P_{a_i,\alpha_i}$.
\item[e)]
It is easy to guess $i_1,\ldots,i_k$ and, based on d), 
verify in polynomial time
that $P^{(i_1)}\cap\ldots\cap P^{(i_k)}\not=\emptyset$.

We establish $\calNP$--hardness by reduction from \textsf{Clique} \cite{NPc}:
Given a graph $G=([n],E)$, choose
$n\cdot(n-1)/2$ pairwise coprime integers $q_{i,\ell}\geq2$, $1\leq i<\ell\leq n$;
for instance $q_{i,\ell}:=p_{i+n\cdot(\ell-1)}$ will do, where
$p_m$ denotes the $m$-th prime number,
found in time polynomial in $n\leq|\langle G\rangle|$ 
(though \emph{not} in $|\langle p_m\rangle|\approx\log m+\loglog m$)
by simple exhaustive search.
Then calculate $a_i:=\prod_{\ell\not=i} q_{i,\ell}$
and observe that $\gcd(a_i,a_j)=q_{i,j}$ for $i\not=j$.
Now start with $\alpha_1:=0$ and 
iteratively for $\ell=2,3,\ldots,n$
determine $\alpha_\ell$ by solving 
the following system of simultaneous congruences:
\begin{equation} \label{e:CRT}
\alpha_\ell \equiv \left\{ \begin{array}{r@{\;\text{ for }\;}l}
\alpha_i \pmod{q_{\ell,i}} & (\ell,i)\in E \\
1+\alpha_i \pmod{q_{\ell,i}} & (\ell,i)\not\in E \end{array}\right.
\quad, \qquad 1\leq i<\ell  \end{equation}
Indeed, as the $q_{\ell,i}$ are pairwise coprime,
the Chinese Remainder Theorem asserts the existence of
a solution---computable in time polynomial in $n$,
regarding that $\alpha_\ell$ can be bounded by $\prod_{i,j} q_{i,j}$
having a polynomial number of bits).
The thus constructed vector $(\alpha_i)_{_i}$ satisfies:
\[ \alpha_i\equiv\alpha_j \pmod{\underbrace{\gcd(a_i,a_j)}_{=q_{i,j}}}
\quad\Leftrightarrow\quad (i,j)\in E \]
because, for $(i,j)\not\in E$, Equation~(\ref{e:CRT})
implies $\alpha_i\equiv\alpha_j\mathbf{+1}\pmod{q_{i,j}}$.

We claim that this mapping $G\mapsto(a_i,\alpha_i:1\leq i\leq n)$
constitutes the desired reduction:
Indeed, according to Equation~(\ref{e:CRT1}),
any sub-collection $P^{(i_1)},\ldots,P^{(i_k)}$ has
non-empty intersection (i.e. a common element $x$)
~iff~ $\alpha_{i_\ell}\equiv\alpha_{i_j} \pmod{\gcd(a_{i_\ell},a_{i_j})}$,
i.e., by our construction, ~iff~ $(i_\ell,i_j)\in E$;
hence cliques of $G$ are in one-to-one correspondence
with subcollections of intersecting arithmetic progressions.
\qed\end{enumerate}\end{proof}

\begin{proof}[Theorem~\ref{t:Aristotle}]
At time $t$, planet $\#i$ appears at angular position $u_i+t\cdot v_i\mod 1$;
and an exact conjunction between $\#i$ and $\#j$ occurs whenever
$u_i+t\cdot v_i=u_j+t\cdot v_j+z$ for some $z\in\IZ$,
that is iff 
\begin{equation} \label{e:Conjunction}
t\;\in\; \Big\{\frac{u_j-u_i}{v_i-v_j}+z\cdot\frac{1}{v_i-v_j}\Big\}
\;=\; P^{(i,j)}:=P_{a_{i,j},\alpha_{i,j}} \quad \text{ where 
$a_{i,j}:=\frac{1}{v_i-v_j}, \alpha_{i,j}:=\frac{u_j-u_i}{v_i-v_j}$} \enspace .
\end{equation}
Therefore, planets $\#1,\ldots,\#n$ attain a conjunction
at some time $t$ ~iff~ $t\in\bigcap_{i=1}^n P^{(1,i)}$.
The existence of such $t$ thus amounts to the non-emptiness of
the joint intersection of arithmetic progressions
and can be decided in the claimed complexity 
according to Lemma~\ref{l:Progression}b+d).
Moreover, Lemma~\ref{l:Progression}a+c+d) shows
how to calculate the smallest $t$.

Concerning $\calNP$--hardness claimed in Item~f), 
we reduce from Lemma~\ref{l:Progression}e):
Given $n$ arithmetic progressions $P^{(i)}=P_{a_i,\alpha_i}$,
let $u_i:=-\alpha_i\cdot a_i$, $v_i:=1/a_i$, and $u_0:=0=:v_0$.
Then conjunctions between $\#0$ and $\#i$ occur exactly
at times $t\in P^{(i)}$; 
and $P^{(i_1)},\ldots,P^{(i_k)}$ meet ~iff~ (and when/where)
$\#0,\#i_1,\ldots,\#i_k$ do.

Approximate conjunction up to $\epsilon$ in time interval $(u,v)$ means:
\[ \exists t\in(u,v) \;\exists z\in\IZ:
  \;(v_2-v_1)t+u_2-u_1+z\in (-\epsilon,+\epsilon) \]
which is equivalent to Claim~b). The boundaries
of the interval $(a,b)\cdot(v_1-v_2)+u_1-u_2
  + (-\epsilon,+\epsilon)$ can be calculated in $\calNC^1$.
\qed\end{proof}
\subsubsection{General Eudoxus/Aristotle; Ptolemy, Copernicus, and Kepler} \label{s:Ptolemy} ~\\
Proceeding from the restricted 2D theory $\Phi'$
to Eudoxus/Aristotle's full $\Phi$
obviously complicates the computational
complexity of the above predictions;
and it seems desirable to make that precise,
e.g. with the help of \cite{Atallah,Gajentaan,Breu}.
Moreover also $\Phi$ in turn had been refined:
\textsc{Ptolemy} introduced additional so-called \emph{epicycles}
and \emph{deferents}, located and rotating on the 
originally earth-centered spheres.
This allowed for a more (parameters to fit in order to yield an)
accurate description of the observed planetary motions.
Copernicus relocated the spheres (and sub-spheres thereon)
to be centered around the sun, rather than earth.
And Kepler replaced them with ellipses in space.
Again, the respective 
increase in complexity is worth-while investigating.

\subsection{Opticks}
There is an abundance of (physical theories giving) explanations
for optical phenomena; cmp. e.g. \emph{The Book of Optics} 
by \textsc{Ibn al-Haytham} (1021) or \textsc{Newton}'s book
providing the title of this section. We are specifically
interested in the progression from 
geometric via Gaussian (taking into account dispersion)
over \textsc{Huygens} and \textsc{Fourier} (diffractive, wave) optics 
to Maxwell's theory of electromagnetism;
and even, in order to describe the various kinds of scattering
observed, to quantum and quantum field theories.
Note that this sequence of optical theories $\Phi_i$ reflects 
their historical succession, but \emph{not} a logical one in the
sense that $\Phi_{i+1}$ `implies' (and hence is computationally
at least as hard as) $\Phi_i$. 

Our purpose is thus to explore more thoroughly the 
computational complexities of these theories.
In fact their computational relations may happen
to be similar, unrelated, or just opposite to
their historical ones!
Consider for example geometric optics versus
Electrodynamics:

\subsubsection{Geometric Optics}
considers light rays as ideal geometric objects, i.e., of infinitesimal
section proceeding instantaneously and straightly until hitting a, say,
mirror. Now depending on the kind of mirrors (straight or
curved, with rational or algebraic parameters) and the availability
of further optical devices (lenses, beam splitters),
\cite{Tygar} has developed a fairly exhaustive taxonomy of
the induced computational complexities of ray tracing
ranging from $\calPSPACE$ to \emph{un}decidable!

\subsubsection{Electrodynamics}
on the other hand treats light as a vector-valued wave
obeying a system of linear partial differential equations
named after \textsc{James Clerk Maxwell}.
Their solution, from given initial conditions, 
is computable, even over real numbers \cite{ZhongWave1}!

\subsection{Quantum Mechanics}
is, since \textsc{Richard P. Feynman}'s famous \textsf{Lectures on Computation}
\cite{Feynman}, of particular interest to the theory of computation and has,
in connection with the work of \textsc{Peter Shor}'s,
initiated Quantum Computation as a now fashionable 
and speculative \cite{Kieu} research topic lacking
a general picture \cite{WDSmithKieu,Myrvold,ZhongSchroedinger}.
Speaking in complexity theoretic terms, the (as usual highly ambiguous)
question raised by the strong CTH (recall Section~\ref{s:Physics})
asks to locate the computational power of \QM
somewhere among (or between) $\calP$, $\calP^{\text{IntegerFactorization}}$, $\calNP$,
and $\Delta_2$.
And it seems worth-while to further explore how
the answer depends on the underlying
Hamiltonians being un-/bounded as indicated
in \mycite{Chapter~3}{PER}? 

In order for a sound and more definite investigation,
our approach suggests to start exploring well-specified
sub- and pre-theories of \QM. These may for instance be
the \textsc{Bohr-Sommerfeld} theory of classical 
electron orbits with integral action-angle conditions.

Another promising direction considers computational
capabilities of, and complexity in, \textsf{Quantum Logic}:

\subsubsection{Quantum Logic} 
arises as an abstraction of the purely algebraic structure
exhibited by the collection of \emph{effect} operators
introduced by \textsc{G.~Ludwig} 
on a Hilbert space (i.e. certain quantum mechanical observables); 
cf. e.g. \cite{Svozil2}.
This discipline has flourished from the comparison with (i.e. systematic
and thorough investigation of similarities and differences to)
Boolean logic. In particular the axioms
satisfied by operations ``$\wedge,\vee,\neg,\leq$'' differ
from the classical case, depending on which quantum logic
one considers.

It seems interesting to devise a theory of computational complexity
similar to that of Boolean circuits 
\cite[\textsc{Sections}~4.3 and 11.4]{Papadimitriou} with classical gates
replaced by quantum \emph{logic} ones. A first important and
non-trivial result has been obtained in \mycite{Section~3}{Hagge}
and may be interpreted as: the satisfiability problem
for quantum logic gates is decidable.


\end{document}

%% file: mymaketitle.tex
\makeatletter
\renewcommand\maketitle{\newpage
  \refstepcounter{chapter}%
  \stepcounter{section}%
  \setcounter{section}{0}%
  \setcounter{subsection}{0}%
  \setcounter{figure}{0}
  \setcounter{table}{0}
  \setcounter{equation}{0}
  \setcounter{footnote}{0}%
  \begingroup
    \parindent=\z@
    \renewcommand\thefootnote{\@fnsymbol\c@footnote}%
    \if@twocolumn
      \ifnum \col@number=\@ne
        \@maketitle
      \else
        \twocolumn[\@maketitle]%
      \fi
    \else
      \newpage
      \global\@topnum\z@   
      \@maketitle
    \fi
    \thispagestyle{empty}\@thanks
    \def\\{\unskip\ \ignorespaces}\def\inst##1{\unskip{}}%
    \def\thanks##1{\unskip{}}\def\fnmsep{\unskip}%
    \instindent=\hsize
    \advance\instindent by-\headlineindent
    \if@runhead
       \if!\the\titlerunning!\else
         \edef\@title{\the\titlerunning}%
       \fi
       \global\setbox\titrun=\hbox{\small\rm\unboldmath\ignorespaces\@title}%
       \ifdim\wd\titrun>\instindent
          \typeout{Title too long for running head. Please supply}%
          \typeout{a shorter form with \string\titlerunning\space prior to
                   \string\maketitle}%
          \global\setbox\titrun=\hbox{\small\rm
          Title Suppressed Due to Excessive Length}%
       \fi
       \xdef\@title{\copy\titrun}%
    \fi
    \if!\the\tocauthor!\relax
      {\def\and{\noexpand\protect\noexpand\and}%
      \protected@xdef\toc@uthor{\@author}}%
    \else
      \def\\{\noexpand\protect\noexpand\newline}%
      \protected@xdef\scratch{\the\tocauthor}%
      \protected@xdef\toc@uthor{\scratch}%
    \fi
    \if@runhead
       \if!\the\authorrunning!
         \value{@inst}=\value{@auth}%
         \setcounter{@auth}{1}%
       \else
         \edef\@author{\the\authorrunning}%
       \fi
       \global\setbox\authrun=\hbox{\small\unboldmath\@author\unskip}%
       \ifdim\wd\authrun>\instindent
          \typeout{Names of authors too long for running head. Please supply}%
          \typeout{a shorter form with \string\authorrunning\space prior to
                   \string\maketitle}%
          \global\setbox\authrun=\hbox{\small\rm
          Authors Suppressed Due to Excessive Length}%
       \fi
       \xdef\@author{\copy\authrun}%
       \markboth{\@author}{\@title}%
     \fi
  \endgroup
  \setcounter{footnote}{\fnnstart}%
  \clearheadinfo}
\makeatother